\documentclass{article} % For LaTeX2e
\usepackage{iclr2025_conference,times}

\usepackage{hyperref}
\usepackage{url}

\title{Mitigating Societal Cognitive Overload in the Age of AI: Challenges and Directions}

\author{Salem Lahlou\\
Mohamed bin Zayed University of Artificial Intelligence\\
\texttt{salem.lahlou@mbzuai.ac.ae} 
}

\iclrfinalcopy % Uncomment for camera-ready version, but NOT for submission.
\begin{document}

\maketitle

\begin{abstract}
Societal cognitive overload, driven by the deluge of information and complexity in the AI age, poses a critical challenge to human well-being and societal resilience. This paper argues that mitigating cognitive overload is not only essential for improving present-day life but also a crucial prerequisite for navigating the potential risks of advanced AI, including existential threats.  We examine how AI exacerbates cognitive overload through various mechanisms, including information proliferation, algorithmic manipulation, automation anxieties, deregulation, and the erosion of meaning.  The paper reframes the AI safety debate to center on cognitive overload, highlighting its role as a bridge between near-term harms and long-term risks.  It concludes by discussing potential institutional adaptations, research directions, and policy considerations that arise from adopting an overload-resilient perspective on human-AI alignment, suggesting pathways for future exploration rather than prescribing definitive solutions.
\end{abstract}

\section{Introduction: AI, Cognitive Overload, and the Looming Governance Crisis}

\subsection{The Age of Societal Cognitive Overload: A Looming Crisis}
We stand at a precipice.  Human societies are increasingly struggling to process the sheer volume and complexity of information in the digital age, a condition dramatically amplified by the rapid proliferation of artificial intelligence (AI).  
While \citet{toffler1970future} foresaw ``future shock'' from accelerating change and \citet{eppler2004concept, bawden2009dark} analyzed individual information overload, Byung-Chul Han, in his critique of neoliberalism and technological domination \citep{han2017psychopolitics}, argues that contemporary society faces a regime of technological domination that exploits and overwhelms the psyche.  This exploitation and overwhelming of the psyche, now dramatically amplified by AI-driven information and complexity, elevates information overload to a systemic crisis: \textbf{societal cognitive overload}.

This is a state where individuals, institutions, and even entire governments are overwhelmed, their capacity to make sound decisions eroded by the sheer cognitive demands of navigating AI-driven systems.

This societal cognitive overload manifests across three critical domains:

\begin{itemize}
    \item \textbf{Informational Domain:} The deluge of AI-generated synthetic media (deepfakes \citep{chesney2019deepfakes,tolosana2020deepfakes}, algorithmically curated filter bubbles \citep{pariser2011filter} and disinformation campaigns \citep{vaccari2020deepfakes}) erode shared epistemic ground \citep{benkler2018network}, making it harder to discern truth from falsehood.
    \item \textbf{Moral Domain:} Societies struggle to define fairness in algorithmic systems \citep{noble2018algorithms} and grapple with fundamental ethical questions about human agency, responsibility, and values in an age of automation \citep{carr2010shallows,turkle2011alone}.
    \item \textbf{Systemic Domain:} As \citet{tainter1988collapse} warned, societies risk collapse when complexity outpaces their capacity to manage it. AI-driven interconnectedness \citep{crawford2021atlas} and the potential erosion of human decision-making skills \citep{kahneman2011thinking} exacerbate this systemic fragility.
\end{itemize}

Critically, this multifaceted cognitive overload not only degrades our capacity to address present-day challenges, across these informational, moral, and systemic domains, but also undermines our ability to grapple with the profound long-term implications of AI. This includes the very real concerns about AI safety and potential existential risks articulated by leading researchers \citep{bostrom2014superintelligence, russell2019human, bengio-x-risk, ord2020precipice}.

Complementing these concerns about abrupt existential risks, recent research by \citet{kulveit2025gradual} highlights the equally concerning, yet often overlooked, threat of ``gradual disempowerment''.  They argue that even incremental advancements in AI, by progressively replacing human labor and cognition across crucial societal systems, can lead to a systemic and potentially irreversible erosion of human influence and control, ultimately precipitating a different pathway to existential catastrophe through the slow and subtle undermining of human agency at a societal scale.

\subsection{AI as a Double-Edged Sword}
AI technologies act as both cause and potential remedy for this overload:
\begin{itemize}
    \item \textbf{Exacerbation}: Algorithmic manipulation traps users in engagement-driven filter bubbles \citep{pariser2011filter}, automation disrupts labor markets \citep{acemoglu2017robots,acemoglu2020automation}, and opaque systems concentrate power in unaccountable platforms \citep{srnicek2017platform,varoufakis2023technofeudalism}.
    \item \textbf{Mitigation}: AI could enhance human cognition through tools for information filtering \citep{malone2018superminds} or decision support, but only if designed to prioritize societal resilience over profit motives \citep{norman2013design}.
\end{itemize}

\subsection{Buried Questions and Institutional Paralysis}
AI forces societies to confront long-deferred questions:
\begin{itemize}
    \item What level of inequality is permissible in economies where AI concentrates wealth \citep{acemoglu2012why}?
    \item How can policymakers govern algorithms when technical complexity overwhelms democratic processes \citep{green2021flaws}?
    \item What existential risks are acceptable in pursuing artificial general intelligence \citep{bostrom2014superintelligence,bengio-x-risk}?
\end{itemize}
These dilemmas remain unresolved because cognitive overload paralyzes institutions. As \citet{green2021flaws} demonstrates, even well-intentioned policies like human oversight of algorithms fail when decision-makers lack the bandwidth to audit complex systems. This creates a \textit{bidirectional misalignment}: overloaded institutions cannot govern AI effectively, while poorly governed AI intensifies societal strain.

\subsection{Paper Outline}
\label{sec:roadmap}
This paper argues that mitigating societal cognitive overload is not merely beneficial, but \textit{essential} for responsible AI development and ensuring alignment with human values, particularly when navigating potential existential risks. To support this argument, we will:
\begin{itemize}
    \item Analyze the key mechanisms by which AI exacerbates societal cognitive overload, spanning informational, economic, and existential dimensions, and highlighting the amplifying roles of deregulation and profit-driven incentives (Section~\ref{sec:double-edged}).
    \item Demonstrate how AI, while exacerbating cognitive overload and potentially paralyzing institutions, paradoxically \textit{forces} us to confront critical ``buried questions'' related to algorithmic fairness, economic inequality, and existential risks that societies can no longer afford to ignore (Section~\ref{sec:buried-questions}).
    \item Conclude by discussing the potential institutional adaptations, research directions, and policy considerations that emerge from an overload-resilient perspective on human-AI alignment, suggesting pathways for future exploration.
\end{itemize}
Our central aim is to reframe the AI alignment challenge through the lens of societal cognitive capacity, arguing that reducing overload is a crucial prerequisite for effective governance, ethical AI development, and ultimately, a safer and more human-compatible AI future.

\section{AI's Double-Edged Sword: Mechanisms of Cognitive Overload}  
\label{sec:double-edged}  

\subsection{Exacerbation: How AI Intensifies Societal Strain}  
\label{sec:exacerbation}  

\subsubsection{Algorithmic Manipulation and Polarization}  
The architecture of AI-driven platforms, optimized for engagement, creates self-reinforcing cycles of polarization. \citet{benkler2018network}'s analysis of the 2016 U.S. election demonstrates how networked propaganda exploits cognitive overload: bots and hyper-partisan media outlets flooded social platforms with disinformation, overwhelming users' capacity to discern truth. As \citet{howard2018algorithms} detail, these actors strategically leverage social media algorithms to amplify their messages, using techniques such as bot networks to artificially inflate engagement and targeted advertising to reach specific demographics with tailored disinformation. This ``cybernetic loop'' of overload and polarization is exacerbated by filter bubbles \citep{pariser2011filter}, where users are trapped in ideological echo chambers. 
For example, the Brexit referendum saw the deployment of AI-powered microtargeting tools, such as those used by Cambridge Analytica \citep{cadwalladr2019cambridge}, which leveraged psychographic profiling to deliver tailored misinformation.  This exploitation of cognitive biases, including confirmation bias, is further amplified under conditions of cognitive overload.  When overloaded, individuals become less capable of engaging the effortful System 2 thinking \citep{kahneman2011thinking} needed for critical evaluation, making them more susceptible to not only initial misinformation, but also to the ``continued influence effect'' \citep{lewandowsky2012misinformation}, where false information persists even after corrections are presented.
This creates a \textit{bidirectional misalignment}: overloaded users are easily swayed by misaligned AI, while these misaligned algorithms further amplify information overload, creating a negative feedback loop. These systems exacerbate societal fragmentation and erode shared reality.

\subsubsection{Automation Anxiety and Economic Fragility}  
\citet{acemoglu2020automation} demonstrates that automation disproportionately displaces low-wage earners in what \citet{autor2003skill} characterize as ``routine-task intensive'' sectors. These sectors, encompassing not only manual labor in manufacturing but also cognitive roles in clerical and administrative work, are particularly vulnerable due to the codifiable and rule-based nature of their tasks, making them readily automatable by AI and robots.
The psychological toll is severe: \citet{case2020deaths} links job displacement to rising ``deaths of despair'' (suicide, substance abuse) in communities hollowed out by automation. 
Cognitive overload further exacerbates this crisis, as workers facing displacement and the daunting prospect of reskilling must navigate complex and often opaque AI-driven job markets \citep{webb2020impact}.  The sheer scale of this reskilling challenge is underscored by OECD studies \citep{arntz2016risk}, which highlight the potential for widespread job displacement across developed economies.  Moreover, policymakers themselves, often facing their own cognitive overload amidst rapid technological change and lobbying pressures \citep{crawford2021atlas, green2021flaws}, may struggle to enact effective and timely responses.
Cognitive overload thus becomes a significant barrier to workers adapting to automation-driven job displacement, hindering their ability to reskill and effectively navigate new, AI-driven job markets. This dynamic contributes to the expansion of what \citet{standing2011precariat} describes as the ``precariat'': a growing class facing precarious employment and economic insecurity.  The cognitive burden experienced by the precariat, coupled with anxieties about future prospects, not only undermines individual well-being but also diminishes societal resilience and the capacity for a constructive and informed public discourse on AI and automation policy, creating a bidirectional misalignment.
This creates a feedback loop where economic precarity erodes societal capacity to govern AI, and unregulated AI deepens inequality.

\subsubsection{Erosion of Human Agency}  
\citet{turkle2011alone} documents how increasing reliance on AI-driven social platforms diminishes empathy and self-reflection, leading to a paradoxical sense of being ``alone together.'' This trend towards shallower online engagement resonates with \citet{carr2010shallows}'s warning about the cognitive consequences of hyperlinked and algorithmically curated content, which fosters ``skimming'' over deep reading.  Indeed, neuroscientific research \citep{small2009your} suggests that habitual internet use may alter brain activity patterns, potentially favoring rapid, shallow information processing at the expense of sustained attention and deep analytical skills.
The cumulative effect of these trends is a society potentially less equipped for deep deliberation, particularly on complex and long-term issues like AI's existential risks. This erosion of human agency is not a mere side-effect, but arguably baked into the design of many platforms.  As \citet{eyal2014hooked} elucidates in ``Hooked'', persuasive design principles are strategically employed to maximize user engagement and habit formation, often at the cost of users' conscious control over their attention and cognitive habits. Furthermore, \citet{dennett2017bacteria}'s concept of ``competence without comprehension'', exemplified by AI systems like based on large language models \citep{brown2020language,openai2023gpt4,guo2025deepseek} generating human-like text without genuine understanding\footnote{The question of whether LLMs genuinely understand language and concepts remains a subject of ongoing philosophical inquiry regarding the nature of understanding and reasoning.}, raises deeper concerns about human agency in relation to increasingly sophisticated AI, echoing \citet{vinge1993coming}'s warnings about a potential ``post-human era'' where human control is fundamentally challenged.
Cognitive overload, exacerbated by reliance on AI-driven platforms, diminishes human self-reflection and critical thinking. This leads to a \textbf{bidirectional misalignment}: humans become less capable of articulating and defending their values in the face of increasingly powerful AI, while AI systems, developed without robust human value input, may drift further from human-compatible goals. This concern about the erosion of human critical thinking skills in the face of increasingly capable AI is echoed in recent commentary, with articles in media outlets like Big Think \citep{pomeroy2025AIcriticalthinking} raising questions about whether over-reliance on AI tools could lead individuals to ``take the convenient route of allowing AI to handle our critical thinking'', rather than preserving and developing this essential cognitive capacity themselves.

\subsubsection{Deregulation, Profit Motives, and Concentration of Power as Cognitive Overload Amplifiers}
The exacerbation of societal cognitive overload by AI is not solely a technological phenomenon; it is deeply intertwined with socio-economic and political factors.  Current trends in deregulation, the dominance of profit motives, and the increasing concentration of power in the hands of a few large tech corporations \citep{varoufakis2023technofeudalism,heikkila2023generativeAIrisks,verdegem2024dismantling} act as significant amplifiers of cognitive overload.  As governments struggle to keep pace with AI's rapid evolution, the cognitive burden of oversight often leads to de facto or explicit deregulation. This ``cognitive offloading'' to the market, as discussed by \citet{braithwaite2000global} in the context of global business regulation, can be particularly problematic in the AI sector.  
Fundamentally, the exacerbation of societal cognitive overload by AI is deeply intertwined with prevailing profit motives and the mechanics of the attention economy \citep{wu2016attention}.  As \citet{lanier2018ten} compellingly argues, the core business model of many dominant tech platforms is not merely about connecting people, but about capturing and relentlessly monetizing user attention. This creates a powerful incentive to design systems that maximize engagement metrics, even if such designs inherently contribute to information overload, algorithmic manipulation, and societal fragmentation, as these often paradoxically increase short-term engagement. This trend is further amplified by the increasing concentration of power and resources within a handful of tech corporations \citep{zuboff2019age, srnicek2017platform}.  As \citet{khan2016amazon} powerfully demonstrates in her analysis of Amazon's ``antitrust paradox'', and as \citet{zuboff2019age} details in her critique of ``surveillance capitalism'', these entities wield immense power to shape the digital information landscape, control vast data resources, and influence public discourse. This concentrated power directly undermines societal resilience to cognitive overload, as the decisions of these corporations, often driven by opaque algorithms and shareholder value maximization, have far-reaching and largely unaccountable impacts on the information environment and the cognitive well-being of billions.
This dynamic reinforces a feedback loop where cognitive overload weakens regulatory capacity, leading to further deregulation and increased corporate power, which in turn intensifies cognitive overload.

\subsubsection{Existential Uncertainty and the Cognitive Load of Meaning-Making}
Beyond the more readily quantifiable forms of cognitive overload, AI introduces a more subtle yet profound cognitive burden: existential uncertainty and a sense of eroding meaning. As AI systems increasingly exhibit capabilities once considered uniquely human—creativity, complex communication, problem-solving—fundamental questions about human identity, purpose, and uniqueness are brought into sharp relief. This growing sense of existential uncertainty is not merely a theoretical concern, but is reflected in public perceptions of AI.  A recent National Public Opinion Poll on the Impact of AI \citep{elon2024publicopinionAI} reveals widespread public anxiety about the societal implications of AI, suggesting a broad societal awareness of the profound and potentially unsettling transformations AI may bring to human life and meaning-making.  This can be understood as a form of ``existential cognitive load'', reflecting the often taxing mental effort and anxiety associated with grappling with these profound questions of meaning and purpose in a world where the boundaries of human and artificial intelligence are becoming increasingly blurred, as explored for example by \citet{yalom2020existential}'s work on existential psychotherapy.
This challenge to anthropocentric worldviews can create a sense of existential unease and disorientation. The democratization of powerful AI tools, like LLMs, makes these existential questions accessible and relevant to a wider population. While this existential questioning can be a catalyst for philosophical reflection and societal evolution, it also undeniably adds to the overall societal cognitive load, as individuals and communities grapple with redefining their place and purpose in a world increasingly co-inhabited by intelligent non-human agents. This challenge to traditional meaning-making deeply resonates with broader philosophical concerns about the ``malaise of modernity'', as \citet{taylor1989sources} describes, where established sources of identity and value are under strain, leaving individuals and societies searching for new foundations.  Furthermore, this existential uncertainty, when combined with the more practical forms of information and task-related cognitive overload, can contribute to a pervasive sense of societal anxiety and a diminished capacity for collective action, potentially exacerbating the decline in social capital and civic engagement highlighted by \citet{putnam2000bowling}.

\subsection{Mitigation: Toward Human-Centered AI Design}
\label{sec:mitigation}

\subsubsection{Context-Aware and Human-Centered Tools for Cognitive Support}
\label{sec:mitigation-tools}
AI's potential for cognitive augmentation hinges on designing tools that genuinely enhance human capabilities without exacerbating overload.  \citet{malone2018superminds} envisions AI as ``superminds'' -- collaborative systems that amplify collective intelligence.  To realize this, AI tools must be context-aware, adaptive to individual cognitive capacities, and prioritize human agency. Realizing this potential for cognitive augmentation, however, is far from straightforward.  The development of truly effective human-centered AI tools for cognitive support faces significant technical and design challenges.  It requires not only advanced AI algorithms but also careful consideration of user interface design, human-computer interaction principles, and ethical implications to ensure these tools genuinely empower users without introducing new forms of overload or manipulation.

\begin{itemize}
    \item Personalized Information Filtering and Sensemaking: AI could offer personalized information filters prioritizing relevance and reducing redundancy, drawing on ``universal usability'' \citep{shneiderman2000universal}. AI-powered summarization tools based on LLMs could reduce cognitive burden of information processing. However, building effective personalized filters that avoid echo chambers (\citet{gomez2015netflix,nguyen2014exploring}) remains a technical hurdle.  Careful UI/UX design, guided by usability principles \citep{nielsen1994usability}, is crucial.

    \item Explainable and Transparent AI for Trust and Calibration:  Mitigation requires Explainable AI (XAI) providing insights into reasoning \citep{doshi2017towards, lipton2018mythos, guidotti2018survey}. XAI helps users understand AI recommendations and calibrate trust.  However, creating usable XAI under cognitive overload remains a challenge (\citet{miller2019explanation,nielsen1994usability}).  XAI should offer concise, relevant, actionable insights, aligning with ``mixed-initiative'' interface design \citep{horvitz1999principles}, where humans and AI collaborate to understand each others' contributions.

    \item Tools for Collective Deliberation and Consensus Building: Platforms like Polis \citep{small2021polis} show AI's potential for large-scale deliberation. AI tools could identify agreement/disagreement, structure discussions, summarize viewpoints.  However, scalability, moderation, UI design for synthesis remain hurdles. Ethical concerns about algorithmic bias and manipulation must be addressed.  These tools aim to enhance ``collective intelligence'' \citep{woolley2010evidence}.
\end{itemize}

However, it's crucial to acknowledge the potential pitfalls.  AI tools themselves can be designed to be addictive \citep{alter2017irresistible}, manipulative, or biased.  Therefore, human-centered design must be guided by ethical principles and focus on empowering users, not further exploiting their cognitive vulnerabilities.  The goal is to create AI that enhances human cognitive capacity and agency, rather than substituting or undermining it.  Context-aware AI tools aim to reduce cognitive overload, thereby enhancing human capacity to understand and interact effectively with AI. This contributes to \textit{bidirectional alignment} by empowering humans to better steer AI and ensuring AI systems are designed to be more human-compatible in cognitively demanding environments.

\subsubsection{Guardrails Against Cognitive Exploitation and Systemic Overload}
\label{sec:mitigation-guardrails}
Technical solutions alone are insufficient.  Mitigating societal cognitive overload also requires robust ``guardrails'': regulatory frameworks and societal norms that protect against cognitive exploitation and systemic risks. These guardrails must address the bidirectional nature of the problem: preventing AI from exacerbating overload, and ensuring institutions are capable of governing AI effectively. However, establishing effective guardrails against cognitive exploitation and systemic overload is a complex undertaking, fraught with practical, ethical, and political challenges.  These guardrails must not only be robust enough to protect against harms but also adaptable to the rapidly evolving AI landscape and sensitive to potential unintended consequences or trade-offs with other societal values, such as innovation and freedom of expression.

\begin{itemize}
    \item Transparency, Accountability, and Auditing Mandates:  Building on the EU AI Act \citep{euAIactEuropeanParliament} and IEEE Ethically Aligned Design \citep{ieee2019ethically}, regulations must mandate transparency for ``high-risk'' AI systems, requiring developers to provide documentation, explainability mechanisms, and undergo independent audits.  This reduces the ``cognitive auditing burden'' on regulators \citep{green2021flaws} by making systems more scrutable.  Furthermore, accountability mechanisms are crucial: assigning clear responsibility for harms caused by AI systems, encouraging responsible development and deployment. However, implementing effective transparency, accountability, and auditing mandates for AI systems is far from trivial.  As \citet{green2021flaws} points out, even well-intentioned human oversight policies can falter due to the sheer cognitive burden of auditing complex algorithms.  Furthermore, ensuring due process and fairness in algorithmic systems, as explored by \citet{citron2014scored}, requires careful consideration of procedural mechanisms and regulatory capacity.  Building this regulatory capacity, including developing AI auditing tools for regulators themselves and fostering interdisciplinary expertise, is a critical challenge.

    \item Attention Economy Reforms and ``Right to Disconnect'' Principles:  The attention economy, driven by engagement-maximizing algorithms \citep{wu2016attention}, directly contributes to cognitive overload and societal polarization.  Reforms could include limiting the capacity of platforms to algorithmically promote harmful content or filter bubbles, and expanding the right to disconnect \citep{muller2020right,syvertsen2020taking,baerten2023right} to protect individuals from constant digital connectivity and work encroachment, fostering cognitive restoration.  However, regulating the attention economy and implementing the right to disconnect involve navigating complex trade-offs.  For example, overly restrictive regulation of algorithmic amplification could raise concerns about freedom of expression and innovation \citep{napoli2019social}.  Similarly, enforcing right to disconnect policies in a globalized and always-on work culture presents practical and cultural challenges.  Policy design in this area requires careful balancing of competing values and a nuanced understanding of the digital media ecosystem.
    \item Labor and Economic Safeguards in the Age of Automation:  To mitigate automation anxiety and economic precarity, which exacerbate cognitive overload, policy interventions are needed:
    \begin{itemize}
        \item Strengthening social safety nets: Universal Basic Income (UBI) or robust unemployment benefits can reduce economic stress and free cognitive resources for adaptation and civic engagement \citep{danaher2019automation}.
        \item Investing in reskilling and lifelong learning:  Preparing the workforce for the changing job market, but also ensuring reskilling programs are cognitively accessible and effective, rather than adding to overload.
        \item Promoting human-centered automation: Encouraging AI development that augments human labor rather than solely replacing it, focusing on tasks that are dull, dangerous, or dirty, while preserving meaningful human work.
    \end{itemize}
    However, even seemingly beneficial policies like UBI are subject to ongoing debate and scrutiny.  As explored in extensive research \citep{bidadanure2019political}, questions remain about the optimal design, funding mechanisms, and potential societal impacts of UBI, including its effects on work motivation and inflation.  Furthermore, ensuring that reskilling programs are truly effective and accessible, and that ``human-centered automation'' is not merely a rhetorical concept but a practical reality, requires sustained effort and careful policy implementation.
\end{itemize}

Responsive regulation \citep{braithwaite2000global}, characterized by iterative, evidence-based rules and stakeholder engagement, is crucial for navigating the rapidly evolving AI landscape.  ``Sandbox'' approaches, like Singapore's \citep{oecd2024regulatory}, can allow for controlled experimentation and impact assessment before widespread deployment, fostering a more adaptive and overload-resilient approach to AI governance.  Ultimately, these guardrails are essential for ensuring that AI benefits society as a whole, rather than exacerbating existing inequalities and cognitive strains. Regulatory guardrails aim to prevent AI from exacerbating cognitive overload and to ensure institutions have the capacity to govern AI effectively. This is essential for \textit{bidirectional alignment}: reducing societal cognitive overload creates a more stable foundation for developing and deploying AI that is truly aligned with human values, while effective governance mechanisms ensure ongoing alignment as AI evolves.

\section{AI as Catalyst: Confronting Buried Societal Questions of Fairness, Inequality, and Risk}
\label{sec:buried-questions}  

While societal cognitive overload presents a significant impediment to addressing complex challenges, the rise of AI paradoxically acts as a catalyst, forcing a long-overdue confrontation with fundamental societal questions that have often remained buried beneath layers of routine and deferred deliberation. In periods of relative societal stability, societies can often postpone grappling with deeply uncomfortable questions about justice, equity, and existential risks.  However, the transformative and disruptive power of AI, coupled with the intensifying pressures of cognitive overload, resurfaces these ``buried questions'' with a new urgency, demanding that we confront them head-on if we are to navigate the AI age responsibly and ethically.

\subsection{Defining Justice in Algorithmic Systems: An AI-Driven Imperative}
Can justice be mathematically defined for AI? \citet{mittelstadt2016ethics} argues fairness metrics conflict, revealing fairness is value-laden. Fairness is inherently value-laden, reflecting diverse and often conflicting human priorities. This complexity, often overlooked, is exposed by codifying fairness for AI.  Cognitive overload paralyzes ethical discourse, yet amplifies the societal imperative to grapple with these questions. Without consensus on values for \textbf{bidirectional alignment}, AI alignment with human interests is impossible, exacerbating societal anxiety and overload. This lack of value consensus undermines AI alignment and erodes public trust in these systems. True justice in the AI age demands participatory design: AI adapting to human values. 

\subsection{Thresholds of Inequality in Automated Economies}  
\citet{acemoglu2012why}'s influential framework in ``Why Nations Fail'' distinguishes sharply between ``inclusive'' and ``extractive'' institutions. \textit{Extractive institutions}, in their analysis, are designed to concentrate power and wealth in the hands of a narrow elite, hindering broad-based economic progress.  In contrast, \textit{inclusive institutions} foster wider participation, protect property rights, and promote competition, creating conditions for innovation and shared prosperity, a distinction that gains critical salience in the context of AI-driven economies. The increasing sophistication and pervasiveness of AI technologies compels us to confront a long-avoided question: What are the acceptable thresholds of economic inequality in societies where AI-driven automation and platform capitalism reshape labor markets and wealth distribution? AI risks entrenching ``extractive'' institutions:  For instance, AI-driven gig economy platforms increasingly employ algorithmic management to maximize efficiency, often at the cost of worker precarity and reduced benefits \citep{srnicek2017platform}. This dynamic contributes to a ``digital precariat'' lacking the resources to advocate for systemic change.  \citet{west2017scale}'s research on urban systems, indicating inequality follows power-law distributions, further highlights AI's potential to amplify existing disparities, potentially leading to a future dominated by a ``cognitive elite'' controlling data and wealth. The looming potential for AI to fundamentally reshape economic structures compels societies to finally grapple with the ethical and political implications of rising inequality and to actively seek pathways toward more inclusive and equitable AI-augmented economies.

\subsection{Existential Risk and Societal Deliberation: AI's Unsettling Provocation}  
\citet{ord2020precipice} estimates a non-negligible risk of human extinction by 2100, with misaligned AI identified as a leading contributor.  While such long-term, low-probability risks can easily be dismissed or deferred in the face of more pressing daily concerns, the sheer scale of potential AI-driven existential threats acts as an unsettling provocation, forcing societies to confront a most fundamental and long-avoided question: What level of existential risk, if any, is ethically and societally acceptable to incur in the pursuit of advanced artificial intelligence? Yet, as we have argued, public deliberation, essential for navigating such complex ethical terrain, is paradoxically paralyzed by ``existential fatigue'': a cognitive overload subtype where individuals disengage from long-term, seemingly remote risks.  The 2017 Asilomar AI Principles \citep{asilomarAIprinciples}, while a valuable expert-driven initiative, notably lacked broader public input, reflecting a critical governance gap in addressing AI's most profound long-term implications.  To bridge this gap and foster more inclusive deliberation, tools like \textit{Deliberative Polls} that use AI to synthesize citizen inputs and create ``cognitive scaffolds'' for informed discourse become increasingly vital. The unprecedented nature of AI-driven existential risks, however remote, serves as a profound catalyst, compelling humanity to develop new modes of long-term societal deliberation and governance capable of grappling with threats that transcend immediate human experience and cognitive biases.

\subsection{Cognitive Overload as a Catalyst for AI Safety Concerns}
The discourse surrounding AI safety, particularly the potential for existential risks, gains critical relevance when viewed through the lens of societal cognitive overload. Leading voices in AI safety, such as Yoshua Bengio \citep{bengio-x-risk}, Stuart Russell \citep{russell2019human}, Nick Bostrom \citep{bostrom2014superintelligence}, and Toby Ord \citep{ord2020precipice}, have articulated concerns about advanced AI systems becoming misaligned with human values, potentially leading to catastrophic outcomes.  However, the capacity of societies to engage thoughtfully with these long-term risks is significantly undermined by cognitive overload.  As argued throughout this paper, cognitive overload impairs decision-making at all levels, erodes social cohesion, and fosters instability. These factors not only exacerbate immediate harms from AI but also create a less resilient and less capable global society for navigating the complex and potentially high-stakes challenges posed by advanced AI, including existential risks.  Therefore, addressing societal cognitive overload is not merely a matter of improving present-day well-being but is also a crucial prerequisite for effectively mitigating potential long-term AI risks.

\section{Conclusion: Reclaiming Cognitive Capacity for Human-AI Alignment}
\label{sec:conclusion}
Societal cognitive overload is not merely an individual burden but a systemic crisis that undermines our capacity to govern complex technologies like AI and to ensure their alignment with human values.  This paper has argued that mitigating cognitive overload is not just a desirable outcome but a \textbf{precondition for achieving meaningful bidirectional human-AI alignment}. While societal cognitive overload presents a formidable challenge, the very act of confronting this crisis, driven by the transformative power of AI, may paradoxically compel societies to finally address long-deferred and fundamental questions about justice, equity, and the future of humanity.

Our analysis of the mechanisms by which AI exacerbates overload points towards several potential pathways for fostering overload-resilient alignment.  These may include the development of independent AI ethics and oversight agencies. Such agencies could play a crucial role in conducting audits of high-risk AI systems to ensure transparency and fairness, developing and promoting ethical guidelines for AI development and deployment, serving as a public resource for information and education on AI risks and benefits, and facilitating public deliberation and stakeholder engagement in AI policy-making. Furthermore, fostering societal cognitive resilience may require investment in centers for digital literacy and cognitive resilience. These centers could be instrumental in developing curricula for schools and lifelong learning programs focused on critical thinking and media literacy, offering training in mindful technology use and overload management, and promoting public awareness campaigns about the attention economy and cognitive well-being. Finally, enhancing participatory AI governance might involve creating and supporting online platforms for informed public deliberation. These platforms could facilitate broader citizen input on AI policy issues, aggregate and synthesize diverse viewpoints to inform policymaking, and provide channels for citizen feedback and oversight of AI systems.

A robust interdisciplinary research agenda becomes essential to deepen our understanding of the cognitive and psychological impacts of AI. Key research areas include quantifying the cognitive and psychological effects of different AI systems, identifying vulnerable populations and specific cognitive vulnerabilities, and developing metrics for measuring societal cognitive overload and resilience. Research should also prioritize developing design principles for human-centered and overload-resilient AI. This includes formulating design guidelines for AI tools that minimize cognitive load and maximize human agency, exploring novel interface designs for human-AI collaboration, and investigating the effectiveness of different XAI techniques in overload contexts. Policy levers for cognitive safeguards could include enhanced regulation of algorithmic transparency and accountability, mandating transparency for high-risk AI systems, requiring audits and impact assessments, and establishing clear lines of accountability. Reforms of the attention economy may be necessary, such as policies aimed at curbing the harms of engagement-maximizing algorithms, regulating algorithmic amplification of harmful content, enacting ``right to disconnect'' legislation, and promoting alternative media models that prioritize quality information. Finally, robust labor and social safety nets in the age of automation are essential to mitigate automation anxiety and economic precarity, which exacerbate cognitive overload. This may involve strengthening social safety nets, investing in reskilling and lifelong learning, and promoting human-centered automation through appropriate policies.

The future of human-AI alignment hinges not just on technical advancements, but on our ability to cultivate a cognitively sustainable and ethically robust relationship with increasingly powerful artificial intelligence.  Only by addressing the challenge of societal cognitive overload can we hope to navigate the complexities of the AI age and safeguard a future where cognitively sustainable human societies flourish alongside increasingly powerful artificial intelligence.

\section*{Acknowledgements}
The author is grateful to Hachem Madmoun, Tom Bosc, Taha Skiredj, Salma André and Harsh Satija for valuable discussions and suggestions.

\bibliography{references}

\begin{thebibliography}{80}
\providecommand{\natexlab}[1]{#1}
\providecommand{\url}[1]{\texttt{#1}}
\expandafter\ifx\csname urlstyle\endcsname\relax
  \providecommand{\doi}[1]{doi: #1}\else
  \providecommand{\doi}{doi: \begingroup \urlstyle{rm}\Url}\fi

\bibitem[Acemoglu \& Restrepo(2017)Acemoglu and Restrepo]{acemoglu2017robots}
Daron Acemoglu and Pascual Restrepo.
\newblock Robots and jobs: Evidence from us labor markets.
\newblock \emph{NBER Working Paper Series}, \penalty0 (23285), 2017.

\bibitem[Acemoglu \& Restrepo(2020)Acemoglu and Restrepo]{acemoglu2020automation}
Daron Acemoglu and Pascual Restrepo.
\newblock Automation and new tasks: How technology displaces and reinstates labor.
\newblock \emph{Journal of Economic Perspectives}, 33\penalty0 (2):\penalty0 3--30, 2020.

\bibitem[Acemoglu \& Robinson(2012)Acemoglu and Robinson]{acemoglu2012why}
Daron Acemoglu and James~A Robinson.
\newblock \emph{Why nations fail: The origins of power, prosperity, and poverty}.
\newblock Crown, 2012.

\bibitem[Alter(2017)]{alter2017irresistible}
Adam Alter.
\newblock \emph{Irresistible: The rise of addictive technology and the business of keeping us hooked}.
\newblock Penguin, 2017.

\bibitem[Arntz(2016)]{arntz2016risk}
M~Arntz.
\newblock The risk of automation for jobs in oecd countries: A comparative analysis.
\newblock 2016.

\bibitem[Autor et~al.(2003)Autor, Levy, and Murnane]{autor2003skill}
David~H Autor, Frank Levy, and Richard~J Murnane.
\newblock The skill content of recent technological change: An empirical exploration.
\newblock \emph{The Quarterly journal of economics}, 118\penalty0 (4):\penalty0 1279--1333, 2003.

\bibitem[Baerten et~al.(2023)Baerten, De~Stefano, and Wouters]{baerten2023right}
Robbe Baerten, Valerio De~Stefano, and Rik Wouters.
\newblock The right to disconnect: An ethical perspective.
\newblock \emph{Available at SSRN 4348227}, 2023.

\bibitem[Bawden \& Robinson(2009)Bawden and Robinson]{bawden2009dark}
David Bawden and Lyn Robinson.
\newblock The dark side of information: overload, anxiety and other paradoxes and pathologies.
\newblock \emph{Journal of information science}, 35\penalty0 (2):\penalty0 180--191, 2009.

\bibitem[Bengio et~al.(2023)Bengio, Hinton, Yao, Song, Abbeel, Harari, Zhang, Xue, Shalev-Shwartz, Hadfield, et~al.]{bengio-x-risk}
Yoshua Bengio, Geoffrey Hinton, Andrew Yao, Dawn Song, Pieter Abbeel, Yuval~Noah Harari, Ya-Qin Zhang, Lan Xue, Shai Shalev-Shwartz, Gillian Hadfield, et~al.
\newblock Managing ai risks in an era of rapid progress.
\newblock \emph{arXiv preprint arXiv:2310.17688}, pp.\ ~18, 2023.

\bibitem[Benkler et~al.(2018)Benkler, Faris, and Roberts]{benkler2018network}
Yochai Benkler, Robert Faris, and Hal Roberts.
\newblock \emph{Network propaganda: Manipulation, disinformation, and radicalization in American politics}.
\newblock Oxford University Press, 2018.

\bibitem[Bidadanure(2019)]{bidadanure2019political}
Juliana~Uhuru Bidadanure.
\newblock The political theory of universal basic income.
\newblock \emph{Annual Review of Political Science}, 22\penalty0 (1):\penalty0 481--501, 2019.

\bibitem[Bostrom(2014)]{bostrom2014superintelligence}
Nick Bostrom.
\newblock \emph{Superintelligence: Paths, dangers, strategies}.
\newblock Oxford University Press, 2014.

\bibitem[Braithwaite \& Drahos(2000)Braithwaite and Drahos]{braithwaite2000global}
John Braithwaite and Peter Drahos.
\newblock \emph{Global business regulation}.
\newblock Cambridge University Press, 2000.

\bibitem[Brown et~al.(2020)Brown, Mann, Ryder, Subbiah, Kaplan, Dhariwal, Neelakantan, Shyam, Sastry, Askell, et~al.]{brown2020language}
Tom Brown, Benjamin Mann, Nick Ryder, Melanie Subbiah, Jared~D Kaplan, Prafulla Dhariwal, Arvind Neelakantan, Pranav Shyam, Girish Sastry, Amanda Askell, et~al.
\newblock Language models are few-shot learners.
\newblock \emph{Advances in neural information processing systems}, 33:\penalty0 1877--1901, 2020.

\bibitem[Cadwalladr(2019)]{cadwalladr2019cambridge}
Carole Cadwalladr.
\newblock Cambridge analytica a year on:‘a lesson in institutional failure’.
\newblock \emph{The Guardian}, 17, 2019.

\bibitem[Carr(2010)]{carr2010shallows}
Nicholas Carr.
\newblock \emph{The shallows: What the Internet is doing to our brains}.
\newblock W. W. Norton \& Company, 2010.

\bibitem[Case \& Deaton(2020)Case and Deaton]{case2020deaths}
Anne Case and Angus Deaton.
\newblock \emph{Deaths of despair and the future of capitalism}.
\newblock Princeton University Press, 2020.

\bibitem[Chesney \& Citron(2019)Chesney and Citron]{chesney2019deepfakes}
Robert Chesney and Danielle Citron.
\newblock Deepfakes and the new disinformation war: The coming age of post-truth geopolitics.
\newblock \emph{Foreign Aff.}, 98:\penalty0 147, 2019.

\bibitem[Citron \& Pasquale(2014)Citron and Pasquale]{citron2014scored}
Danielle~Keats Citron and Frank Pasquale.
\newblock The scored society: Due process for automated predictions.
\newblock \emph{Wash. L. Rev.}, 89:\penalty0 1, 2014.

\bibitem[Crawford(2021)]{crawford2021atlas}
Kate Crawford.
\newblock \emph{Atlas of AI: Power, politics, and the planetary costs of artificial intelligence}.
\newblock Yale University Press, 2021.

\bibitem[Danaher(2019)]{danaher2019automation}
John Danaher.
\newblock \emph{Automation and utopia: Human flourishing in a world without work}.
\newblock Harvard University Press, 2019.

\bibitem[Dennett(2017)]{dennett2017bacteria}
Daniel~C Dennett.
\newblock \emph{From bacteria to Bach and back: The evolution of minds}.
\newblock W. W. Norton \& Company, 2017.

\bibitem[Doshi-Velez \& Kim(2017)Doshi-Velez and Kim]{doshi2017towards}
Finale Doshi-Velez and Been Kim.
\newblock Towards a rigorous science of interpretable machine learning.
\newblock \emph{arXiv preprint arXiv:1702.08608}, 2017.

\bibitem[Eppler \& Mengis(2004)Eppler and Mengis]{eppler2004concept}
Martin~J Eppler and Jeanne Mengis.
\newblock The concept of information overload: A review of literature from organization science, accounting, marketing, mis, and related disciplines.
\newblock \emph{The information society}, 20\penalty0 (5):\penalty0 325--344, 2004.

\bibitem[{European Parliament}(2023)]{euAIactEuropeanParliament}
{European Parliament}.
\newblock {EU AI Act: first regulation on artificial intelligence}, June 2023.
\newblock URL \url{https://www.europarl.europa.eu/topics/en/article/20230601STO93804/eu-ai-act-first-regulation-on-artificial-intelligence}.

\bibitem[Eyal(2014)]{eyal2014hooked}
Nir Eyal.
\newblock \emph{Hooked: How to build habit-forming products}.
\newblock Penguin, 2014.

\bibitem[{Future of Life Institute (FLI)}(2017)]{asilomarAIprinciples}
{Future of Life Institute (FLI)}.
\newblock Asilomar ai principles, August 2017.
\newblock URL \url{https://futureoflife.org/open-letter/ai-principles/}.
\newblock Coordinated by FLI and developed at the Beneficial AI 2017 conference.

\bibitem[Gomez-Uribe \& Hunt(2015)Gomez-Uribe and Hunt]{gomez2015netflix}
Carlos~A Gomez-Uribe and Neil Hunt.
\newblock The netflix recommender system: Algorithms, business value, and innovation.
\newblock \emph{ACM Transactions on Management Information Systems (TMIS)}, 6\penalty0 (4):\penalty0 1--19, 2015.

\bibitem[Green(2021)]{green2021flaws}
Ben Green.
\newblock The flaws of policies requiring human oversight of government algorithms.
\newblock \emph{Computer Law and Security Review}, 2021.
\newblock \doi{10.1016/j.clsr.2022.105681}.

\bibitem[Guidotti et~al.(2018)Guidotti, Monreale, Ruggieri, Turini, Giannotti, and Pedreschi]{guidotti2018survey}
Riccardo Guidotti, Anna Monreale, Salvatore Ruggieri, Franco Turini, Fosca Giannotti, and Dino Pedreschi.
\newblock A survey of methods for explaining black box models.
\newblock \emph{ACM computing surveys (CSUR)}, 51\penalty0 (5):\penalty0 1--42, 2018.

\bibitem[Guo et~al.(2025)Guo, Yang, Zhang, Song, Zhang, Xu, Zhu, Ma, Wang, Bi, et~al.]{guo2025deepseek}
Daya Guo, Dejian Yang, Haowei Zhang, Junxiao Song, Ruoyu Zhang, Runxin Xu, Qihao Zhu, Shirong Ma, Peiyi Wang, Xiao Bi, et~al.
\newblock Deepseek-r1: Incentivizing reasoning capability in llms via reinforcement learning.
\newblock \emph{arXiv preprint arXiv:2501.12948}, 2025.

\bibitem[Han(2017)]{han2017psychopolitics}
Byung-Chul Han.
\newblock \emph{Psychopolitics: Neoliberalism and new technologies of power}.
\newblock Verso Books, 2017.

\bibitem[Heikkil{\"a}(2023)]{heikkila2023generativeAIrisks}
Melissa Heikkil{\"a}.
\newblock Generative ai risks concentrating big tech’s power. here’s how to stop it.
\newblock \emph{MIT Technology Review}, April 2023.
\newblock URL \url{https://www.technologyreview.com/2023/04/18/1071727/generative-ai-risks-concentrating-big-techs-power-heres-how-to-stop-it/}.

\bibitem[Horvitz(1999)]{horvitz1999principles}
Eric Horvitz.
\newblock Principles of mixed-initiative user interfaces.
\newblock \emph{Proceedings of the SIGCHI conference on Human factors in computing systems}, pp.\  159--166, 1999.

\bibitem[Howard et~al.(2018)Howard, Woolley, and Calo]{howard2018algorithms}
Philip~N Howard, Samuel Woolley, and Ryan Calo.
\newblock Algorithms, bots, and political communication in the us 2016 election: The challenge of automated political communication for election law and administration.
\newblock \emph{Journal of information technology \& politics}, 15\penalty0 (2):\penalty0 81--93, 2018.

\bibitem[{Imagining the Digital Future Center, Elon University}(2024)]{elon2024publicopinionAI}
{Imagining the Digital Future Center, Elon University}.
\newblock The national public opinion poll on the impact of ai.
\newblock Technical report, Imagining the Digital Future Center, Elon University, February 2024.
\newblock URL \url{https://imaginingthedigitalfuture.org/reports-and-publications/the-impact-of-artificial-intelligence-by-2040/the-national-public-opinion-poll/}.

\bibitem[Kahneman(2011)]{kahneman2011thinking}
Daniel Kahneman.
\newblock \emph{Thinking, fast and slow}.
\newblock Farrar, Straus and Giroux, 2011.

\bibitem[Khan(2016)]{khan2016amazon}
Lina~M Khan.
\newblock Amazon's antitrust paradox.
\newblock \emph{Yale lJ}, 126:\penalty0 710, 2016.

\bibitem[Kulveit et~al.(2025)Kulveit, Douglas, Ammann, Turan, Krueger, and Duvenaud]{kulveit2025gradual}
Jan Kulveit, Raymond Douglas, Nora Ammann, Deger Turan, David Krueger, and David Duvenaud.
\newblock Gradual disempowerment: Systemic existential risks from incremental ai development.
\newblock \emph{arXiv preprint arXiv: 2501.16946}, 2025.

\bibitem[Lanier(2018)]{lanier2018ten}
Jaron Lanier.
\newblock \emph{Ten arguments for deleting your social media accounts right now}.
\newblock Random House, 2018.

\bibitem[Lewandowsky et~al.(2012)Lewandowsky, Ecker, Seifert, Schwarz, and Cook]{lewandowsky2012misinformation}
Stephan Lewandowsky, Ullrich~KH Ecker, Colleen~M Seifert, Norbert Schwarz, and John Cook.
\newblock Misinformation and its correction: Continued influence and successful debiasing.
\newblock \emph{Psychological science in the public interest}, 13\penalty0 (3):\penalty0 106--131, 2012.

\bibitem[Lipton(2018)]{lipton2018mythos}
Zachary~C Lipton.
\newblock The mythos of model interpretability: In machine learning, the concept of interpretability is both important and slippery.
\newblock \emph{Queue}, 16\penalty0 (3):\penalty0 31--57, 2018.

\bibitem[Malone(2018)]{malone2018superminds}
Thomas~W Malone.
\newblock \emph{Superminds: The surprising power of people and computers thinking together}.
\newblock Little, Brown Spark, 2018.

\bibitem[Miller(2019)]{miller2019explanation}
Tim Miller.
\newblock Explanation in artificial intelligence: Insights from the social sciences.
\newblock \emph{Artificial intelligence}, 267:\penalty0 1--38, 2019.

\bibitem[Mittelstadt et~al.(2016)Mittelstadt, Allo, Taddeo, Wachter, and Floridi]{mittelstadt2016ethics}
Brent~Daniel Mittelstadt, Patrick Allo, Mariarosaria Taddeo, Sandra Wachter, and Luciano Floridi.
\newblock The ethics of algorithms: Mapping the debate.
\newblock \emph{Big Data \& Society}, 3\penalty0 (2):\penalty0 2053951716679679, 2016.

\bibitem[M{\"u}ller(2020)]{muller2020right}
Klaus M{\"u}ller.
\newblock The right to disconnect.
\newblock \emph{European Parliamentary Research Service Blog}, 9, 2020.

\bibitem[Napoli(2019)]{napoli2019social}
Philip Napoli.
\newblock \emph{Social media and the public interest: Media regulation in the disinformation age}.
\newblock Columbia university press, 2019.

\bibitem[Nguyen et~al.(2014)Nguyen, Hui, Harper, Terveen, and Konstan]{nguyen2014exploring}
Tien~T Nguyen, Pik-Mai Hui, F~Maxwell Harper, Loren Terveen, and Joseph~A Konstan.
\newblock Exploring the filter bubble: the effect of using recommender systems on content diversity.
\newblock In \emph{Proceedings of the 23rd international conference on World wide web}, pp.\  677--686, 2014.

\bibitem[Nielsen(1994)]{nielsen1994usability}
Jakob Nielsen.
\newblock \emph{Usability engineering}.
\newblock Morgan Kaufmann, 1994.

\bibitem[Noble(2018)]{noble2018algorithms}
Safiya~Umoja Noble.
\newblock \emph{Algorithms of oppression: How search engines reinforce racism}.
\newblock NYU Press, 2018.

\bibitem[Norman(2013)]{norman2013design}
Donald~A Norman.
\newblock \emph{The design of everyday things: Revised and expanded edition}.
\newblock Basic books, 2013.

\bibitem[OECD(2024)]{oecd2024regulatory}
OECD.
\newblock Regulatory experimentation: Moving ahead on the agile regulatory governance agenda.
\newblock Technical report, 2024.

\bibitem[OpenAI(2023)]{openai2023gpt4}
OpenAI.
\newblock Gpt-4 technical report, 2023.
\newblock URL \url{https://arxiv.org/abs/2303.08774}.

\bibitem[Ord(2020)]{ord2020precipice}
Toby Ord.
\newblock \emph{The precipice: Existential risk and the future of humanity}.
\newblock Hachette Books, 2020.

\bibitem[Pariser(2011)]{pariser2011filter}
Eli Pariser.
\newblock \emph{The filter bubble: What the Internet is hiding from you}.
\newblock Penguin Press, 2011.

\bibitem[Pomeroy(2025)]{pomeroy2025AIcriticalthinking}
Ross Pomeroy.
\newblock Is ai eroding our critical thinking?
\newblock \emph{Big Think}, January 2025.
\newblock URL \url{https://bigthink.com/thinking/artificial-intelligence-critical-thinking/}.

\bibitem[Putnam(2000)]{putnam2000bowling}
Robert~D Putnam.
\newblock \emph{Bowling alone: The collapse and revival of American community}.
\newblock Simon and Schuster, 2000.

\bibitem[Russell(2019)]{russell2019human}
Stuart Russell.
\newblock \emph{Human compatible: Artificial intelligence and the problem of control}.
\newblock Viking, 2019.

\bibitem[Shneiderman(2000)]{shneiderman2000universal}
Ben Shneiderman.
\newblock Universal usability.
\newblock \emph{Communications of the ACM}, 43\penalty0 (5):\penalty0 84--91, 2000.

\bibitem[Small et~al.(2021)Small, Bjorkegren, Erkkil{\"a}, Shaw, and Megill]{small2021polis}
Christopher Small, Michael Bjorkegren, Timo Erkkil{\"a}, Lynette Shaw, and Colin Megill.
\newblock Polis: Scaling deliberation by mapping high dimensional opinion spaces.
\newblock \emph{Recerca: revista de pensament i an{\`a}lisi}, 26\penalty0 (2), 2021.

\bibitem[Small et~al.(2009)Small, Moody, Siddarth, and Bookheimer]{small2009your}
Gary~W Small, Teena~D Moody, Prabha Siddarth, and Susan~Y Bookheimer.
\newblock Your brain on google: patterns of cerebral activation during internet searching.
\newblock \emph{The American Journal of Geriatric Psychiatry}, 17\penalty0 (2):\penalty0 116--126, 2009.

\bibitem[Srnicek(2017)]{srnicek2017platform}
Nick Srnicek.
\newblock \emph{Platform capitalism}.
\newblock Polity, 2017.

\bibitem[Standing(2011)]{standing2011precariat}
Guy Standing.
\newblock The precariat: The new dangerous class.
\newblock \emph{Bloomsbury academic}, 2011.

\bibitem[Syvertsen(2020)]{syvertsen2020taking}
Trude Syvertsen.
\newblock Taking back control! the right to disconnect and mobile technology.
\newblock \emph{New Technology, Work and Employment}, 35\penalty0 (3):\penalty0 249--254, 2020.

\bibitem[Tainter(1988)]{tainter1988collapse}
Joseph~A Tainter.
\newblock \emph{The collapse of complex societies}.
\newblock Cambridge university press, 1988.

\bibitem[Taylor et~al.(1989)]{taylor1989sources}
Charles Taylor et~al.
\newblock \emph{Sources of the self: The making of the modern identity}, volume 1989.
\newblock Harvard University Press Cambridge, MA, 1989.

\bibitem[{The IEEE Global Initiative on Ethics of Autonomous and Intelligent Systems}(2019)]{ieee2019ethically}
{The IEEE Global Initiative on Ethics of Autonomous and Intelligent Systems}.
\newblock Ethically aligned design: A vision for prioritizing human well-being with autonomous and intelligent systems.
\newblock Technical report, IEEE, Piscataway, NJ, 2019.
\newblock URL \url{https://standards.ieee.org/content/ieee-standards/en/industry-connections/ec/autonomous-systems.html}.

\bibitem[Toffler(1970)]{toffler1970future}
Alvin Toffler.
\newblock Future shock, 1970.
\newblock \emph{Sydney. Pan}, 1970.

\bibitem[Tolosana et~al.(2020)Tolosana, Vera-Rodriguez, Fierrez, Morales, and Ortega-Garcia]{tolosana2020deepfakes}
Ruben Tolosana, Ruben Vera-Rodriguez, Julian Fierrez, Aythami Morales, and Javier Ortega-Garcia.
\newblock Deepfakes and beyond: A survey of face manipulation and fake detection.
\newblock \emph{Information Fusion}, 64:\penalty0 131--148, 2020.

\bibitem[Turkle(2011)]{turkle2011alone}
Sherry Turkle.
\newblock \emph{Alone together: Why we expect more from technology and less from each other}.
\newblock Basic Books, 2011.

\bibitem[Vaccari \& Chadwick(2020)Vaccari and Chadwick]{vaccari2020deepfakes}
Cristian Vaccari and Andrew Chadwick.
\newblock Deepfakes and disinformation: Exploring the impact of synthetic political video on deception, uncertainty, and trust in news.
\newblock \emph{Social media+ society}, 6\penalty0 (1):\penalty0 2056305120903408, 2020.

\bibitem[Varoufakis(2023)]{varoufakis2023technofeudalism}
Yanis Varoufakis.
\newblock \emph{Technofeudalism: What killed capitalism}.
\newblock Random House, 2023.

\bibitem[Verdegem(2024)]{verdegem2024dismantling}
Pieter Verdegem.
\newblock Dismantling ai capitalism: the commons as an alternative to the power concentration of big tech.
\newblock \emph{AI \& society}, 39\penalty0 (2):\penalty0 727--737, 2024.

\bibitem[Vinge(1993)]{vinge1993coming}
Vernor Vinge.
\newblock Coming technological singularity: How to survive in the post-human era.
\newblock \emph{Vision-21: Interdisciplinary Science and Engineering in the Era of Cyberspace}, 1993.

\bibitem[Webb(2020)]{webb2020impact}
Michael Webb.
\newblock The impact of artificial intelligence on the labor market.
\newblock \emph{Available at SSRN 3482150}, 2020.

\bibitem[West(2017)]{west2017scale}
Geoffrey West.
\newblock \emph{Scale: The universal laws of growth, innovation, sustainability, and the pace of life in organisms, cities, economies, and companies}.
\newblock Penguin Press, 2017.

\bibitem[Woolley et~al.(2010)Woolley, Chabris, Pentland, Hashmi, and Malone]{woolley2010evidence}
Anita~Williams Woolley, Christopher~F Chabris, Alex Pentland, Nada Hashmi, and Thomas~W Malone.
\newblock Evidence for a collective intelligence factor in the performance of human groups.
\newblock \emph{science}, 330\penalty0 (6004):\penalty0 686--688, 2010.

\bibitem[Wu(2016)]{wu2016attention}
Tim Wu.
\newblock \emph{The attention merchants: The epic scramble to get inside our heads}.
\newblock Knopf, 2016.

\bibitem[Yalom(2020)]{yalom2020existential}
Irvin~D Yalom.
\newblock \emph{Existential psychotherapy}.
\newblock Hachette UK, 2020.

\bibitem[Zuboff(2019)]{zuboff2019age}
Shoshana Zuboff.
\newblock \emph{The age of surveillance capitalism: The fight for a human future at the new frontier of power}.
\newblock PublicAffairs, 2019.

\end{thebibliography}
\bibliographystyle{iclr2025_conference}

\end{document}